\documentclass{ws-ijmpd}
\usepackage[super,compress]{cite}
\begin{document}

\markboth{Jose Luis Bl\'azquez Salcedo}
%{Extremal rotating black holes in Einstein-Maxwell-Chern-Simons theory: radially excited solutions and non-uniqueness}
{Extremal rotating black holes in Einstein-Maxwell-Chern-Simons theory%: radially excited solutions and non-uniqueness
}

\catchline{}{}{}{}{}
%
%%%%%%%%%%%%%%%%%%%%%%%%%%%%%%%%%%%%%%%%%%%%%%%%%%%%%%%%%%%%%%%%%%%%

\title{\uppercase{Extremal rotating black holes in Einstein-Maxwell-Chern-Simons theory: radially excited solutions and non-uniqueness}
}

\author{\uppercase{Jose Luis Bl\'azquez-Salcedo}
}
\address{Institut f\"ur Physik, Universit\"at Oldenburg Postfach 2503, D-26111 Oldenburg, Germany
\\
jose.blazquez.salcedo@uni-oldenburg.de}

\maketitle

\begin{history}
\received{Day Month Year}
\revised{Day Month Year}
\end{history}

\begin{abstract}
We study 5-dimensional black holes in Einstein-Maxwell-Chern-Simons theory with free Chern-Simons
coupling parameter. We consider an event horizon with spherical topology, and both angular momenta
of equal magnitude. In particular, we study extremal black holes, which can be used to obtain the
boundary of the domain of existence. Above a critical value of the Chern-Simons coupling constant we
find non-static extremal solutions with vanishing angular momentum. These solutions form a sequence
which can be labeled by the node number of the magnetic $U(1)$ potential or the inertial dragging. As the node number
increases, their mass converges to the mass of the extremal Reissner-Nordstr\"om solution. The
near-horizon geometry of the solutions of this sequence is the same. In general not all near-horizon solutions are found as
global solutions, and we show non-uniqueness between extremal solutions and non-extremal ones.
\end{abstract}


\ccode{PACS numbers: 04.40.Nr, 04.20.Jb, 04.50.-h, 04.70.-s}

\section{Introduction}	

The study of black holes in higher-dimensional theories of gravity is important for our understanding of gravity and high energy physics. Developments such as the statistical counting of black hole microstates  for the five-dimensional extremal Reissner-Nordstr\"om solution \cite{Strominger199699} have increased the interest in the study of black holes in different higher-dimensional generalizations of General Relativity.

The higher-dimensional generalization of the Kerr solution was discovered by Myers and Perry (MP solution) \cite{Myers1986304}. Since then, other analytical solutions have been found in different modified theories of gravity motivated by supergravity and string theory, typically coupling gravity to other fields, like for example an electromagnetic field and a dilaton \cite{Kunz200695}. 

In general, stationary black holes in $D=2N+1$ dimensions possess $N$ independent
angular momenta $J_{i}$, which are associated with the $N$ orthogonal planes of rotation \cite{Kunz2006362} and the rank of the rotation group $SO(D-1)$. In this paper we are interested in 5-dimensional black holes with all $N=2$ angular momenta of equal-magnitude. In this case the space-time has cohomogeneity-1 symmetry.

We consider a generalization of the 5-dimensional Einstein-Maxwell theory (EM), by supplementing the action with a Chern-Simons term (CS)
\begin{equation} \label{EMCSac}
I= \frac{1}{16\pi G_5} \int d^5x\biggl[ 
\sqrt{-g} \, (R -% F^{2}
F_{\mu \nu} F^{\mu \nu}
)
-
\frac{2\lambda}{3\sqrt{3}}\,\varepsilon^{\mu\nu\alpha\beta\gamma}A_{\mu}F_{\nu\alpha}F_{\beta\gamma} 
 \biggr ].
\end{equation}
Note that $R$ is the curvature scalar,
$A_\mu $ is the gauge potential with field strength tensor $ F_{\mu \nu}
= \partial_\mu A_\nu -\partial_\nu A_\mu $, $\lambda$ is the CS coupling parameter and
$G_5$ is Newton's constant in five dimensions (we fix the normalization so that $16\pi \, G_5=1$ in the following). Here we are interested in asymptotically flat space-times.

%, and minimal 5-dimensional gauged supergravity is found for $\lambda=\lambda_{\rm SG}=1$. 
An analytical solution is known for a particular value of the CS coupling parameter $\lambda=\lambda_{SG} \equiv 1$ , corresponding to minimally gauged supergravity \cite{Breckenridge1996423,Breckenridge199793,Cvetic2004273,PhysRevLett.95.161301}. In this case, a subset of cohomogeneity-1 solutions corresponds to the Breckenridge-Myers-Peet-Vafa solutions (BMPV) \cite{Breckenridge199793}. Although these black holes have null horizon angular velocity, their angular momentum is not zero, and they are not static solutions.   

In this paper we will study the generalization of these rotating and charged EMCS black holes \cite{PhysRevLett.95.161301} to general values of the CS coupling. In particular we will consider the case in which the CS
coefficient is increased beyond the critical value of $2\lambda_{SG}$. For these values of the CS coupling, the topologically spherical black holes are no longer uniquely determined by the value of the global charges \cite{PhysRevLett.96.081101,PhysRevLett.112.011101}. 

In order to generate the black hole solutions, we will make use of both numerical methods (for the calculation of the global solutions) and analytical methods (the near-horizon formalism). Let us begin by presenting some details about the theory, the Ansatz we use, and how the charges are calculated. 

\section{Einstein-Maxwell-Chern-Simons theory and properties of the black holes}

The field equations of the action (\ref{EMCSac}) are the 
Einstein equations 
\begin{equation}
\label{Einstein_equation}
G_{\mu\nu}=\frac{1}{2} F_{\mu\rho} {F_\nu}^\rho 
  - \frac{1}{8} g_{\mu \nu} F_{\rho \sigma} F^{\rho \sigma},
\end{equation}
and the 
Maxwell-Chern-Simons equations
\begin{equation}
\label{Maxwell_equation}
\nabla_{\nu} F^{\mu\nu} + \frac{\lambda}{2\sqrt{3}}\varepsilon^{\mu\nu\alpha\beta\gamma}F_{\nu\alpha}F_{\beta\gamma}=0.
\end{equation}

The stationary black holes with both angular momenta of equal magnitude,
$|J_{(1)}|=|J_{(2)}|=J$, and horizon of spherical topology, can be parametrized with the following Ansatz for the metric
\cite{Kunz2006362}
\begin{eqnarray}
\label{metric}
&&ds^2 = -f(r) dt^2 + \frac{m(r)}{f(r)}(dr^2 + r^2 d\theta^2)  + \frac{n(r)}{f(r)}r^2 \sin^2\theta \left( d \varphi_1 -\frac{\omega(r)}{r}dt
\right)^2 \nonumber \\  && + \frac{n(r)}{f(r)}r^2 \cos^2\theta \left( d \varphi_2
  -\frac{\omega(r)}{r}dt \right)^2 + \frac{m(r)-n(r)}{f(r)}r^2 \sin^2\theta \cos^2\theta(d \varphi_1  -d \varphi_2)^2,
\end{eqnarray}
where $\theta \in [0,\pi/2]$, $\varphi_1 \in [0,2\pi]$ and $\varphi_2 \in [0,2\pi]$.
For the gauge potential the Ansatz is 
\begin{equation}
A_\mu dx^\mu  = a_0(r) dt + a_{\varphi}(r) (\sin^2 \theta d\varphi_1+\cos^2 \theta d\varphi_2).
\end{equation}
The unknown metric and gauge potential functions depend only on the radial coordinate $r$, which we will assume to be isotropic.

We will characterize the black holes by their global charges and horizon charges. Using the Komar expressions we can obtain the total mass $M$ and the angular momentum $J$
\begin{equation}
M = -  \frac{3}{2} \int_{S_{\infty}^{3}} \alpha  
\ , \label{Kmass} \end{equation}
\begin{equation}
J =   \int_{S_{\infty}^{3}} \beta_{(k)} 
\ , \label{Kang} \end{equation}
where $\alpha_{\mu_1 \mu_2 \mu_3} \equiv \epsilon_{\mu_1 \mu_2 \mu_3
  \rho \sigma} \nabla^\rho \xi^\sigma$,
and
$\beta_{ (k) \mu_1 \mu_2 \mu_3} \equiv \epsilon_{\mu_1 \mu_2 \mu_3
  \rho \sigma} \nabla^\rho \eta_{(k)}^\sigma$.
	Note that $\xi \equiv \partial_t$, $\eta_{(1)} \equiv \partial_{\varphi_1}$ and $\eta_{(2)} \equiv \partial_{\varphi_2}$ are the Killing vectors of these stationary black hole space-times.

The electric charge $Q$ is given by
\begin{equation}
Q= - \frac{1}{2} \int_{S_{\infty}^{3}} \tilde F 
\ , \label{charge} \end{equation}
where
${\tilde F}_{\mu_1 \mu_2 \mu_3} \equiv  
  \epsilon_{\mu_1 \mu_2 \mu_3 \rho \sigma} F^{\rho \sigma}$.

In isotropic coordinates the black hole horizon is at $r=r_{\rm H}$, where we impose the Killing horizon conditions, and the horizon rotates with angular velocity $\Omega_{\rm H}$. 
Regularity of the Killing horizon \cite{Hollands2007} imposes boundary conditions on the metric functions, in particular
\begin{eqnarray}
\Omega_{\rm H} = \frac{\omega(r_{\rm H})}{r_{\rm H}}.
\end{eqnarray}

With this particular choice of coordinates, extremal black holes are obtained when $r_{\rm H}=0$. In this case we have that the degenerated Killing horizon rotates with angular velocity 
\begin{eqnarray}
\Omega_{\rm H} = \omega'(r_{\rm H}).
%\frac{d\omega}{dr}(r_{\rm H}).
\end{eqnarray}

The area $A_{\rm H}$ of the horizon will be of interest to us and it is given by the expression
\begin{equation}
A_{\rm H}=\int_{{\cal H}} \sqrt{|g^{(3)}|}=r_{\rm H}^{3} A(S^{3}) \lim_{r \to r_{\rm H}}
 \sqrt{\frac{m^{2} n}{f^{3}}} \label{hor_area} . \end{equation}
The area is related to the entropy by $S = 4\pi A_{\rm H}$.
The
horizon angular momenta
$J_{{\rm H} (k)}$ are
\begin{equation}
J_{{\rm H} (k)} =   \int_{{\cal H}} \beta_{(k)} \ 
 , \label{Hang} \end{equation}
where ${\cal H}$ represents the surface of the horizon.
Note that when we have equal-magnitude angular momenta, we also have
$|J_{{\rm H} (k)}| =J_{\rm H}$, $k=1, 2$.

\section{Near-horizon geometry}

As commented in the introduction, we can also gain some information about the properties 
of extremal EMCS black holes if we study near-horizon solutions
in the entropy function formalism
\cite{1126-6708-2005-09-038,1126-6708-2006-10-058,1126-6708-2007-11-049}.
This formalism allows us to obtain analytic expressions for the horizon charges of extremal black holes, such as the horizon area and horizon angular momentum. Some global charges such as the electric charge and the angular momentum can also be obtained. Here the CS term of the action needs to be treated carefully \cite{0264-9381-24-20-009}.

All known examples of extremal black holes with event horizon of spherical topology have a near-horizon geometry with $AdS_2 \times S^{D-2}$ symmetry. Hence we can make use of an Ansatz for the near-horizon geometry assuming these isometries (and also in our particular case, cohomogeneity-1)
\cite{KunduriLRR2013}
\begin{eqnarray}
\label{metric_ansatz}
ds^2 &=& v_1(dr^2/r^2-r^2dt^2) + v_2[4d\theta^2+\sin^22\theta(d\varphi_2-d\varphi_1)^2]
\\ 
\nonumber
&+&v_2\eta[d\varphi_1+d\varphi_2+\cos{2\theta}(d\varphi_2-d\varphi_1)-\alpha r dt]^2.
\end{eqnarray}
We have shifted the radial coordinate $r\rightarrow r - r_{\rm H}$ so the horizon is at $r=0$.

For the gauge potential we write
\begin{eqnarray}
\label{gv_ansatz}
A &=& -(\rho + p\alpha)rdt + 2p(\sin^2\theta d\varphi_1 
+ \cos^2\theta d\varphi_2).
\end{eqnarray}
The Ansatz parameters $v_1$, $v_2$, $\eta$, $\alpha$, $\rho$ and $p$ 
satisfy some constraints which can be obtained using the near-horizon formalism 
\cite{1126-6708-2005-09-038,1126-6708-2006-10-058,1126-6708-2007-11-049}, or alternatively solving the field equations.
The resulting relations are
\begin{eqnarray}
\label{relgen}
&&v_2 = v_1, \nonumber\\
&&\eta v_1 = -\frac{4}{3}\frac{(\rho-p+p\alpha)(\rho+p+p\alpha)}{\alpha^2-1}, \nonumber\\
&&v_1 = \frac{2}{3}\frac{\alpha^4p^2-p^2+2\alpha^3\rho p-4\rho p \alpha + \alpha^2\rho^2 - 2\rho^2}{\alpha^2-1}, \nonumber
\\
&&3\alpha v_2^{5/2}\sqrt\eta\rho + 3\alpha^2v_2^{5/2}\sqrt\eta p - 4\lambda\sqrt 3 p v_1 v_2 \rho - 4 \lambda \sqrt 3 p^2 v_1 v_2 \alpha \nonumber \\
&& - 3p v_1^2\sqrt v_2 \sqrt \eta = 0.
\end{eqnarray}

These algebraic relations leave two undetermined parameters, related to the angular momentum $J$ 
and the electric charge $Q$ of the extremal black hole.

In the presence of a Chern-Simons term, the calculation of the angular momentum $J$ 
and the electric charge $Q$ in the near-horizon formalism is somewhat subtle. The correct way to calculate these charges is by making use of the Noether charges 
\cite{0264-9381-24-20-009}. If we follow this procedure we arrive at the following expressions for the charges.
The angular momentum $J$ can be written as
\begin{eqnarray}
J 
= 64\pi^2\frac{v_2^{3/2}}{v_1}\sqrt \eta p(\rho+p\alpha) + 16\pi^2\frac{v_2^{5/2}}{v_1}\eta^{3/2}\alpha - \frac{256}{9}\sqrt 3 \pi^2p^3\lambda,
\end{eqnarray}
and the electric charge $Q$ as
\begin{eqnarray}
Q 
= -64\pi^2\frac{v_2^{3/2}}{v_1}\sqrt \eta (\rho+p\alpha) + \frac{128\pi^2\sqrt 3}{3}\lambda p^2.
\end{eqnarray}

We can obtain some horizon charges using the near-horizon formalism. 
The Komar formula can be used to calculate the horizon angular momentum yielding
\begin{eqnarray}
J_{\rm H} 
=  16\pi^2\frac{v_2^{5/2}}{v_1}\eta^{3/2}\alpha,
\end{eqnarray}
and the horizon area can be written as
\begin{eqnarray}
A_{\rm H} 
= 16\pi^2v_2^{3/2}\sqrt \eta.
\end{eqnarray}

\section{Results for $\lambda>2$}

As we said in the introduction, in this paper we will discuss the properties of the black holes when $\lambda>2$. In particular we choose $\lambda=5$ and fix the electric charge to $|Q|=1$. The features discussed in the following are found for $\lambda>2$, and any other particular value of $|Q|$.

The global solutions are found by
solving the EMCS equations together with the usual boundary conditions obtained from regularity at the horizon and asymptotic flatness. We have used COLSYS, a collocation method for boundary-value
ordinary differential equations, with an adaptive
mesh selection procedure \cite{Ascher:1981}. We solve the differential equations using a compactified radial coordinate, and we employ typically a mesh
of $10^3-10^4$ points, achieving a relative accuracy of $10^{-8}$. 

\begin{figure}[pb]
\centerline{\psfig{file=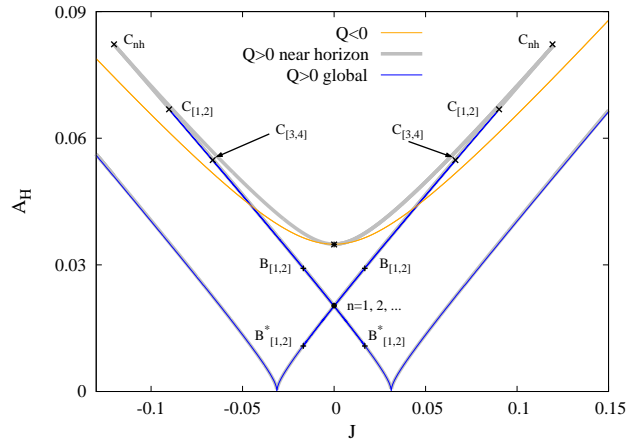,width=6cm,angle=-90}}
\vspace*{8pt}
\caption{Area $A_{H}$ versus angular momentum $J$. We plot extremal solutions for $\lambda=5$ and $|Q|=1$. For positives values of the electric charge, we compare the near-horizon solutions (fat line) with the global solutions (thin line). We mark the cusps $C_{[...]}$ with $\times$. We also mark the branching points $B_{[...]}$ with $+$. Finally we mark the static solution with $\ast$, and the non-static $J=0$ solutions with $\bullet$.   \label{plot_Ah_J}}
\end{figure}

The addition of the CS term implies that the theory is no longer invariant under a change of sign of the electric charge, if the CS coupling is fixed. Hence we have to discuss separately the case with positive charge and negative charge. 

In Figure \ref{plot_Ah_J} we plot the area $A_{\rm H}$ versus the angular momentum $J$ for near-horizon solutions and global solutions ($\lambda=5$). Let us start discussing the $Q=-1$ solutions. In this case we obtain that every solution obtained in the near-horizon formalism corresponds to a global solution, i.e. all the near-horizon solutions are realized globally. Black holes with negative electric charge are always regular, and charged non-static solutions can be connected continuously with the charged static solution by continuously decreasing the angular momentum.

The properties of black holes with positive electric charge $Q=1$ are completely different. Let us start discussing the near-horizon solutions. If we start with some non-static solution (for example with $J=0.15$) and then we decrease the angular momentum, we will eventually reach a singular solution with zero area. The near-horizon solutions then continue as the angular momentum decreases to zero with increasing area. But note that this first $J=0$ near-horizon solution is not the static solution (which has larger area). Instead, this near-horizon solution has $\alpha\neq 0$, and hence is a non-static solution of the near-horizon formalism. The solutions continue with negative values of the angular momentum until we reach a cusp $C_{nh}$, where the curve bends backwards. Finally, the solutions continue down to end at the static solution (marked with $*$ in the figure), where they meet with the symmetric set of solutions ($J \rightarrow -J$).

Now consider the global solutions. Let us start again at some non-static solution with $J=0.15$. Again we decrease the angular momentum and we reach the zero area solution. At this side of the curve, all the near-horizon solutions are realized globally (the same happens with the symmetric solutions with negative values of the angular momentum).

The global solutions continue if we decrease the angular momentum with increasing area. Eventually we reach a non-static $J=0$ solution (we can label this solution $n=1$). The curve continues with negative values of the angular momentum up to a cusp, $C_{[1,2]}$. It is important to note that this cusp is always different from $C_{nh}$. The curve bends backwards and we reach another non-static $J=0$ solution ($n=2$). Eventually, the curve ends at $B^*_{[1,2]}$ at some $J>0$.

Since the cusp $C_{[1,2]}$ is different from the cusp $C_{nh}$, we can see clearly that for $Q>0$ not all the near-horizon solutions are realized globally, meaning that there are solutions from the near-horizon formalism which do not correspond to global solutions of the EMCS theory.

\begin{figure}[pb]
\centerline{\psfig{file=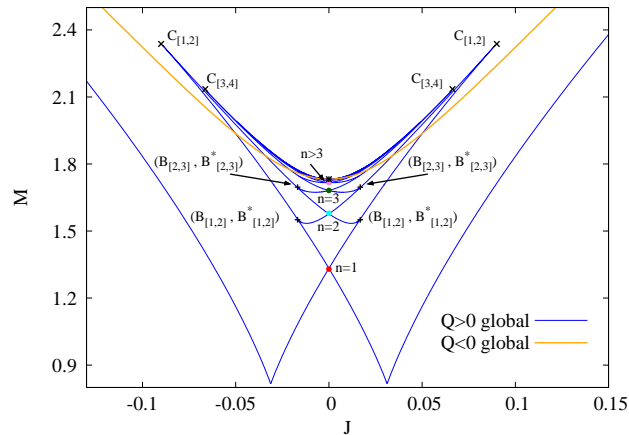,width=6cm,angle=-90}}
\vspace*{8pt}
\caption{Mass $M$ versus angular momentum $J$. We plot global extremal solutions for $\lambda=5$ and $|Q|=1$.  We mark the cusps $C_{[...]}$ with $\times$. We also mark the branching points $B_{[...]}$ with $+$. Finally we mark the static solution with $\ast$, and the non-static $J=0$ solutions with $\bullet$ of different colors. \label{plot_M_J}}
\end{figure}

Nevertheless, Figure \ref{plot_Ah_J} is not appropriate to represent the complicated branch structure of these black holes. It is better to plot global charges, as the mass. We do so in Figure \ref{plot_M_J}, where we plot the mass $M$ versus angular momentum $J$ for the same set of parameters. While the $Q<0$ solutions are only found in a simple branch which ends at the static solution, $Q>0$ black holes present a very complicated bifurcation pattern. 

Minimal mass is reached at the singular solution with vanishing area. Suppose we start at the singular solution with $J>0$ and then we decrease the angular momentum. Eventually we reach $J=0$, with a non-static solution with mass below the static solution ($n=1$). If we now make $J$ negative we can generate a whole branch of extremal black holes, which eventually reach $J=0$, with a different non-static solution ($n=2$) with mass still below the static mass, but above the previous non-static solution. This branch of black holes disappears at a $J>0$ solution $B^*_{[1,2]}$, when the branch touches the symmetrical branch with $J \rightarrow -J$ (we call this a bifurcation point). This solution, $B^*_{[1,2]}$, has the same angular momentum, electric charge and mass as those of a solution of the opposite branch $B_{[1,2]}$. But $B_{[1,2]}$ and $B^*_{[1,2]}$ have different near-horizon geometries. In fact, if we go back to Figure \ref{plot_Ah_J}, we can see that $B_{[1,2]}$ has lower area than $B^*_{[1,2]}$. This is an example of non-uniqueness between two extremal solutions, $B_{[1,2]}$ and $B^*_{[1,2]}$, which share all the global charges. These solutions can be distinguished only by their horizon properties. 

\begin{figure}[pt]
\centerline{\psfig{file=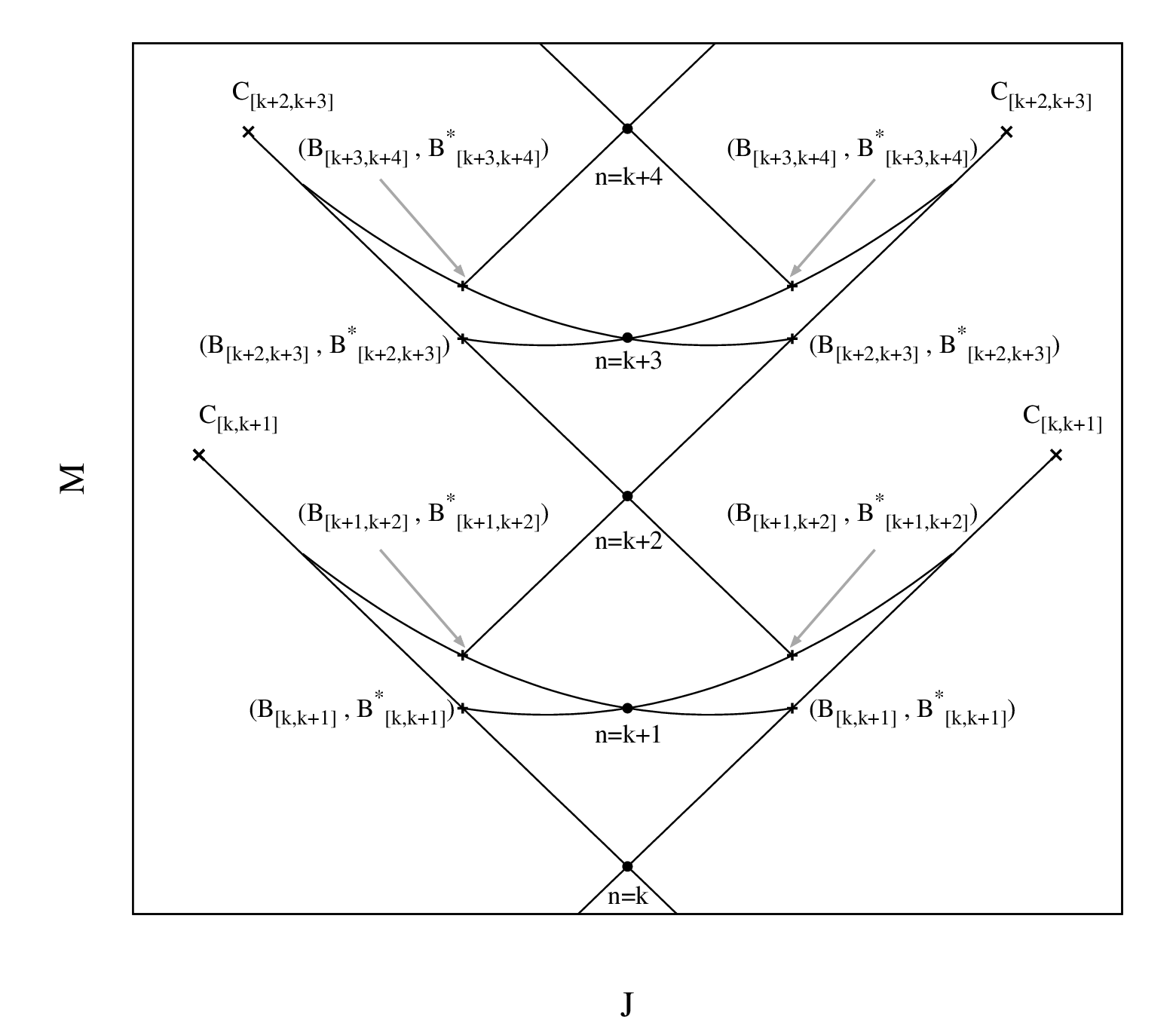,angle=-90,scale=0.3}}
\vspace*{8pt}
\caption{Schematic representation of the branch structure around
the extremal rotating $J\neq 0$ solutions between $n=k$ and $n=k+4$. We mark the cusps $C_{[...]}$ with $\times$. We also mark the branching points $B_{[...]}$ with $+$. Finally we mark the static solution with $\ast$, and the non-static $J=0$ solutions with $\bullet$. \label{plot_M_J_sq}}
\end{figure}

This structure of cusps and bifurcation points repeats an infinite number of times, and the bifurcation pattern gets closer and closer to the static solution. At the bifurcation points we have uniqueness violation, and branches can end or start. In Figure \ref{plot_M_J_sq} we show a schematic representation of the branch structure where the mass is plotted versus the angular momentum. The width of the branches becomes smaller as we go up the bifurcation pattern.

The bifurcation pattern contains an infinite number of non-static $J=0$ solutions. We can label each one of these black holes with an integer number $n=1, 2 ,3 ...$. Note that these black holes have all the same near-horizon geometry as can be seen in Figure \ref{plot_Ah_J}. Hence note that this is an example of infinite non-uniqueness of the near-horizon solution. The non-static $J=0$ solutions can be distinguished by their global charges, like the mass, which increases with $n$ and approaches the mass of the static solution.

\begin{figure}[pt]
\centerline{\psfig{file=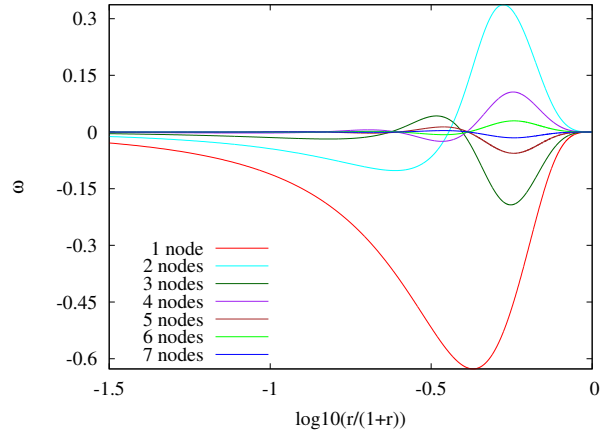,width=6cm,angle=-90}}
\vspace*{8pt}
\caption{The inertial dragging $\omega$ as a function of the compactified radial coordinate. We represent up to 7 non-static $J=0$ solutions, although numerically we can obtain solutions with more than $30$ nodes. \label{plot_omega_x}}
\end{figure}

In fact, the integer number $n$ is related to the node number of the magnetic gauge potential $a_{\varphi}$ and the metric function $\omega$, as can be seen in Figure \ref{plot_omega_x}. Hence these solutions constitute a set of radially excited extremal
solutions.

Interestingly, the static solution is rotationally isolated from the non-static solutions, in the sense that it is not possible to reach the static solution by continuously decreasing the angular momentum or the angular velocity of a non-static solution. We can see this easily if we go back to Figure \ref{plot_Ah_J}. There we can see that all the near-horizon solutions connecting the static solution with the cusp $C_{nh}$ are not realized globally, and the near-horizon geometry of the non-static $J=0$ solutions is different from the near-horizon geometry of the static one.

\begin{figure}[pb]
\centerline{\psfig{file=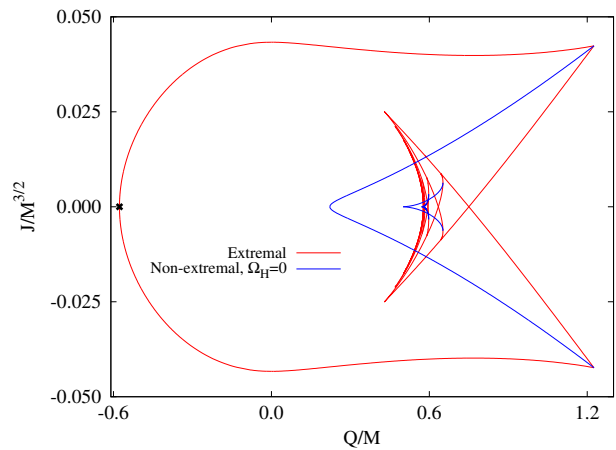,width=6cm,angle=-90}}
\vspace*{8pt}
\caption{Scaled angular momentum $J/M^{3/2}$
versus scaled electric charge $Q/M$ for extremal black holes and $\lambda=5$. We also include non-extremal solutions with $\Omega_{\rm H}=0$ \label{plot_q_j}}
\end{figure}

Finally, in Figure \ref{plot_q_j} we plot the domain of existence of the black holes we are considering. Extremal black holes of fixed mass are expected to be the boundary of the domain of existence. To represent this boundary, we plot in Figure \ref{plot_q_j} the scaled angular momentum $J/M^{3/2}$ versus the scaled electric charge $Q/M$ for $\lambda=5$. But the general features of this plot are found for any $\lambda>2$. 

Every non-extremal solution is found in the inner region delimited by the most external curves of the plot. Note the asymmetry between negative and positive electric charge. The bifurcation pattern is only found for positive $Q$ when $\lambda>2$. Note that the bifurcation pattern is formed by lines inside of the domain of existence. This means that the bifurcation pattern exhibits another type of non-uniqueness. There are non-extremal solutions with the same global charges as the extremal solutions of the bifurcation pattern. As an example of this, we include in the figure non-extremal solutions with null angular velocity but finite angular momentum. The resulting non-extremal curves intersect the bifurcation pattern, meaning that there is uniqueness violation at the crossing points.

\section{Conclusions}

Let us briefly summarize the main results of this paper.

We have considered rotating black holes in Einstein-Maxwell-Chern-Simons theory in 5 dimensions, which are asymptotically flat, and
possess a spherical event horizon topology. We have restricted to the cohomogeneity-1 case, when both angular momenta have equal
magnitude. 

We have used numerical methods to calculate the global solutions, but we have also made use of the near-horizon formalism in order to obtain analytical solutions describing the region near the horizon. Analytical expressions relating the horizon area and angular momentum to the electric charge and total angular momentum have been obtained.

We have centered our discussion on the $\lambda>2$ case. We have found that beyond this critical value of the CS coupling $\lambda$ and for a fixed value of $Q>0$, an infinite sequence of branches
of extremal black holes arises, with very interesting features. We have found that not all the near-horizon solutions are realized globally and alternatively, one near-horizon solution can be related to more than one global solution.

An infinite sequence of radially excited non-static $J=0$ solutions is present in this bifurcation pattern. All the solutions of this sequence of black holes have the same near-horizon geometry, and they can be labeled by an integer number related to the number of nodes of the $a_{\varphi}$ and $\omega$ functions.

We have found that uniqueness is violated between extremal solutions: global charges do not uniquely characterize extremal black holes at the branching points, and some horizon charge should be used in order to distinguish between them.

Even more, we have found that uniqueness can also be violated between extremal and non-extremal solutions, since the complicated bifurcation pattern of extremal black holes is inside the boundary of the domain of existence.

\section*{Acknowledgments}

JLBS would like to thank Jutta Kunz, Francisco Navarro-L\'erida and Eugen Radu for valuable discussions, comments and suggestions on this paper. This work was supported by the Spanish Ministerio de Ciencia e Innovaci\'on, research project FIS2011-28013, and by the DFG, in particular, the DFG Research Training Group 1620 “Models of Gravity.”

\section{References}

\bibliographystyle{ws-ijmpd}
\bibliography{blazquez_aveiro_proc_3}

\end{document}